\documentclass[11pt]{article}
\usepackage{amsmath}

\usepackage{times}

\usepackage{moreverb}  %

\begin{document} \sloppy
\pagenumbering{roman}
\setcounter{page}{1}
\thispagestyle{empty}
\begin{center}
\vspace*{-1in}
Argonne National Laboratory \\
9700 South Cass Avenue\\
Argonne, IL 60439

\vspace{.2in}
\rule{1.5in}{.01in}\\ [1ex]
ANL/MCS-TM-249 \\
\rule{1.5in}{.01in}

\vspace{1in}
{\Large\bf MACE 2.0 Reference Manual and Guide}

\vspace{.2in}
by \\ [3ex]

{\Large\it William McCune}\\
\thispagestyle{empty}

\vspace{1.5in}
Mathematics and Computer Science Division

\bigskip

Technical Memorandum No. 249

\vspace{1in}
May 2001
\end{center}

\vfill
\noindent
This work was supported by the Mathematical,
Information, and Computational Sciences Division subprogram of the
Office of Advanced Scientific Computing Research, U.S. Department of
Energy, under Contract W-31-109-Eng-38.
\newpage
\pagenumbering{roman}
\setcounter{page}{2}
\noindent
Argonne National Laboratory, with facilities in the states of Illinois
and Idaho, is owned by the United States Government and operated by The
University of Chicago under the provisions of a contract with the
Department of Energy.

\vspace{2in}

\begin{center}
{\bf DISCLAIMER}
\end{center}

\noindent
This
report was prepared as an account of work sponsored by an agency of the United States Government.  Neither the United States
Government nor any agency thereof, nor The University of Chicago, nor any of
their employees or officers, makes any warranty, express or implied, or assumes any legal liability or
responsibility for the accuracy, completeness, or usefulness of any
information, apparatus, product, or process disclosed, or represents that its use
would not infringe privately-owned rights.
Reference herein to any specific commercial product, process, or service by trade name,
trademark, manufacturer, or otherwise, does not necessarily constitute
or imply its endorsement, recommendation, or favoring by the United
States Government or any agency thereof.  The views and opinions of document
authors expressed herein do not necessarily state or reflect those of the
United States Government or any agency thereof, Argonne National
Laboratory, or The University of Chicago.

\newpage
  \pagestyle{plain}
  \tableofcontents
\newpage

\pagenumbering{arabic}
\setcounter{page}{1}
\title{MACE 2.0 Reference Manual and Guide}

\author{\emph{William McCune}}

\date{}

\maketitle

\addcontentsline{toc}{section}{Abstract}
\begin{abstract}
MACE is a program that searches for finite models
of first-order statements.  The statement to be modeled is
first translated to clauses, then to relational
clauses; finally for the given domain size, the ground
instances are constructed.  A Davis-Putnam-Loveland-Logeman
procedure decides the propositional problem, and any models found
are translated to first-order models.
MACE is a useful complement to the theorem prover Otter,
with Otter searching for proofs and MACE looking for
countermodels.
\end{abstract}

\section{Introduction}

MACE (Models And CounterExamples) is a program that searches
for small finite models of first-order statements.
It is frequently used along with our first-order theorem prover
Otter~\cite{otter3,otter-web}, with Otter searching for
proofs and MACE looking for countermodels.  The two programs
accept almost the same language, so the same input file can
usually be used for both programs.

MACE has been used for many applications including quasigroup existence
problems~\cite{dp-quasi}, ortholattice problems~\cite{ortholattice}, and
lattice and Boolean algebra problems~\cite{wm-rp:monograph,monograph-web}.
Other successful programs that look for finite models of first-order
statements are SEM~\cite{sem} and FINDER~\cite{finder}.
A related class of programs can produce finite models when the
search for a refutation fails.  Examples of these are
SATCHMO~\cite{satchmo} and MGTP~\cite{mgtp}.

At its core, MACE has a Davis-Putnam-Loveland-Logeman propositional
decision procedure named ANLDP.  ANLDP can be used directly to decide
propositional (SAT) problems given in conjunctive normal form
(see Section~\ref{anldp}).
Section~\ref{history} gives differences between
MACE 2.0 and previous versions.
The MACE Web site is \texttt{http://www.mcs.anl.gov/AR/mace}.

\section{A Little Motivation}

Say you've just invented group theory by writing down
the following three axioms,
\begin{align*}
e * x & = x\\
g(x) * x & = e\\
(x * y) * z & = x * (y * z)
\end{align*}
and you are wondering whether all groups are commutative.
You prepare the following input file, named \texttt{group.in}.
\begin{center}
\begin{small}
\begin{boxedverbatim}
set(auto).
list(usable).
  e * x = x.                  %
  g(x) * x = e.               %
  (x * y) * z = x * (y * z).  %
  a * b != b * a.             %
end_of_list.
\end{boxedverbatim}
\end{small}
\end{center}
\noindent
Now you give the input file to Otter to search for a proof
(actually a refutation).

\begin{small}
\begin{verbatim}
    %
\end{verbatim}
\end{small}
\noindent and to MACE to look for a countermodel (actually a model) of size 4,
as follows.

\begin{small}
\begin{verbatim}
    %
\end{verbatim}
\end{small}
\noindent Both programs fail immediately.  But looking at the Otter
output makes you suspect that not all groups are commutative,
so you go forward, looking for larger countermodels.
The command

\begin{small}
\begin{verbatim}
    %
\end{verbatim}
\end{small}
\noindent succeeds, and the output file contains the following
noncommutative group of order 6.

\begin{small}
\begin{verbatim}
    ======================= Model #1 at 1.13 seconds:
    e: 0
    a: 1
    b: 2
    *:   | 0 1 2 3 4 5
       --+------------
       0 | 0 1 2 3 4 5
       1 | 1 0 3 2 5 4
       2 | 2 4 0 5 1 3
       3 | 3 5 1 4 0 2
       4 | 4 2 5 0 3 1
       5 | 5 3 4 1 2 0
       
    g:     0 1 2 3 4 5
       ---------------
           0 1 2 4 3 5
\end{verbatim}
\end{small}
\noindent Hmmm, very interesting:  I wonder what happens if we add $x*x=e$
to our theory.

\section{How to Tell MACE What to Do}

Three kinds of input determine how MACE works.
First, the clauses or formulas in the input file specify
the theory for which you seek a model.
Second, special commands in the input file put
constraints on the models.
Third, command-line options give general constraints
on the search.

\subsection{The Formulas}

MACE reads the input (from stdin) and takes formulas and clauses from
the lists \texttt{usable}, \texttt{sos}, \texttt{demodulators}, and 
\texttt{passive} as its basic theory.\footnote
{
One can argue that the hot list should also be considered as
part of the basic theory, because Otter uses the hot list to
make inferences.  MACE ignores the hot list, however, because hot list
clauses almost always occur also in usable or sos, and MACE
suffers if it gets duplicate clauses.  I suppose MACE could
get around this by doing a subsumption check.
}
Like Otter, MACE immediately transforms any first-order formulas
to clauses.

MACE is a bit more restrictive than Otter in the language it accepts,
and it interprets some symbols differently.  See Section~\ref{language}.

\subsection{Constraints in the Input}

Constraints are specified in an optional
list \texttt{mace\_constraints} in the input file.\footnote
{
Previous versions of MACE used the \texttt{passive} list for constraints.
}
(If you give Otter an input file containing a \texttt{mace\_constraints} list,
Otter ignores it.)
Two kinds of constraint are accepted:
assignments for the models and
properties of relations or functions.
Here is an example list that shows all of the types of constraint.

\begin{small}
\begin{verbatim}
    list(mace_constraints).
      assign(e, 0).                   %
      assign(g(2), 1).                %
      assign(3*4, 2).                 %
      assign(P(1), T).                %
      assign(Q(0,3), F).              %
      property(same(_,_), equality).
      property(lt(_,_), order).
      property(g(_), bijection).
      property(_*_, quasigroup).
    end_of_list.
\end{verbatim}
\end{small}
The assignments simply give function values or relation values
for particular members of the domain.\footnote
{
Why not place assignments in with the clauses that specify the theory?
This can be done, but such assignments might not make sense if
the input is also being used for Otter.
}
Members of the domain
are always named $0, 1, \ldots, n-1$, where $n$ is the domain size.
The Boolean constants (relation values) are named \texttt{T} and
\texttt{F}.  Note that assigning values to constants can also
be done with the \texttt{-c} command-line option (see the next subsection).
The following properties of function and relation symbols can be specified
in the \texttt{mace\_constraints} list.
\begin{description}
\item{\texttt{equality}}\\
This applies to binary relation symbols.
It is necessary only if a nonstandard equality symbol is being used,
because any binary relation recognized by Otter as an equality symbol
is also recognized by MACE as an equality symbol.
See Section~\ref{syntax}.
\item{\texttt{order}}\\
This applies to binary relation symbols.
It is necessary only if a nonstandard order symbol is being used.
MACE (but not Otter) automatically recognizes binary $<$ as
an order relation.
The ``order'' is the obvious order on
the members of the domain: $0<1<\ldots<n-1$.
See the example input files
\texttt{ordered\_semi.in} and \texttt{cd.in}
included in the MACE distribution package.
\item{\texttt{bijection}}\\
This applies to unary function symbols.
The list of function values is a permutation of the domain.
\item{\texttt{quasigroup}}\\
This applies to binary function symbols.
If you write down the table for a finite quasigroup,
each row and each column is a permutation of the
domain.
\end{description}

\subsection{Command-Line Options}

\newenvironment{clo}%
        {\begin{list}{}{\renewcommand{\makelabel}[1]{\texttt{##1}\hfil}%
                \setlength{\labelwidth}{1.25cm}%
                \setlength{\leftmargin}{2cm}}}%
        {\end{list}}

\begin{clo}
\item[-n $n$]
This gives the starting domain size for the search.
The default value is 2.  If you also give an \texttt{-N} option,
MACE will iterate domain sizes up through the \texttt{-N}
value.
Otherwise, MACE will search only for the \texttt{-n} value.
For example,
\begin{center}
\begin{tabular}{ccc}
\hline
Options &\hspace*{.5cm}& Search \\
\hline
\texttt{-n4}     & & 4 \\
\texttt{-N6}     & & 2,3,4,5,6 \\
\texttt{-n4 -N6} & & 4,5,6 \\
\hline
\end{tabular}
\end{center}
\item[-N $n$]
This gives the ending domain size for the search.
The default is the value of the \texttt{-n} option.
\item[-c]
This says that constants in the input should be assigned
unique elements of the domain.  If the number of constants
in the input is greater than the domain size $n$, the first
$n$ constants are given values, and the rest are unconstrained.
This is a useful option because it eliminates lots
of isomorphism from the search.  But it can block all models,
especially when used with other constraints.
\item[-p]
(Lower case.) This option tells MACE to print models in a nice
tabular form as they are found.  This format is meant
for human consumption.
\item[-P]
(Upper case.) This option tells MACE to print models in an easily parsable
form.  This format has an Otter-like syntax and can be read by
most Prolog systems.
\item[-I]
This option tells MACE to print models in IVY form.
This format is a Lisp S-expression and is meant to be read
by IVY~\cite{ivy}, our proof and model checker.
\item[-m $n$]
This tells MACE to stop after finding $n$ models.  The default is 1.
\item[-t $n$]
This tells MACE to stop after about $n$ seconds.  The default is unlimited.
\emph{MACE ignores any \texttt{assign(max\_seconds, n)} commands
that might be in the input file.  Such commands are used by Otter only.}
\item[-k $n$]
This tells MACE to stop if it tries to allocate more than $n$
kilobytes ofmemory. The default is 48000 (about 48 megabytes).
\emph{MACE ignores any \texttt{assign(max\_mem, n)} commands
that might be in the input file.  Such commands are used by Otter only.}
\item[-x]
This is a special-purpose constraint designed to reduce
isomorphism in quasigroup problems.
It applies only to binary function \texttt{f}.
See~\cite{dp-quasi}.
\item[-h]
This tells MACE to print a summary of these command-line options.
\end{clo}

\section{Language Accepted by MACE} \label{language}

MACE accepts nearly the same input as Otter.
First we list the main differences from Otter; then we give a short
review of Otter's language.

\subsection{Differences from Otter's Language}

\begin{enumerate}
\item MACE does not accept function symbols with arity
greater than three or relation symbols with arity greater
than four.
\item MACE does not allow symbols with different arities,
for example, \texttt{f(f,x)}.
\item MACE does not allow a symbol to be used as both a
relation symbol and a function symbol.
\item
MACE ignores answer literals.  In fact, MACE removes all
answer literals before it starts looking for models.
\item
The natural numbers \texttt{0,1,2,\ldots} are ordinary
constants to Otter, but they have special meanings to MACE.
In particular, MACE interprets them as elements of the domain.
If you ask MACE to look for a model of size $n$, and there are constants
$\geq n$ in the input, MACE will get confused and quit with an error message.
\item
On the other hand,
the evaluable (``dollar'') functions and relations, for example \texttt{\$SUM}
and \texttt{\$LT}, have special meanings to Otter, but they are treated by
MACE as ordinary symbols.  As a result, an input file containing
evaluable symbols can produce both a refutation with Otter and a model
with MACE.  Here is an example.
\end{enumerate}
\begin{center}
\begin{small}
\begin{boxedverbatim}
set(hyper_res).
list(sos).
  -P(x) | P($SUM(x,x)).
  P(1).
  -P(2).
end_of_list.
\end{boxedverbatim}
\end{small}
\end{center}

\subsection{A Quick Review of the Language Otter Accepts} \label{syntax}

See the Otter manual~\cite{otter3} for a thorough description of the language.

\paragraph{Clauses vs.~Formulas.}  You can use either clauses
or formulas.  (Most people use clauses.  If you use formulas,
they are immediately translated to clauses.)  Here are some
corresponding examples.

\begin{small}
\begin{verbatim}
    list(usable).           %
      -P(x) | -Q(x) | R(x).
      -P(x) | -Q(x) | S(x).
      f(e,x) = x.
      f(g(x),x) = e.
    end_of_list.

    formula_list(usable).   %
      all x (P(x) & Q(x) -> R(x) & S(x)).
      exists e ((all x (f(e,x) = x)) &
                (all x exists y (f(y,x) = e))).
    end_of_list.
\end{verbatim}
\end{small}

\paragraph{Variables vs.~Constants in Clauses.}
Clauses do not have explicit quantifiers, so we need a rule to
distinguish variables from constants.  The default rule is that
symbols starting with \texttt{u} through \texttt{z} are variables.  If
the command \texttt{set(prolog\_style\_variables)} is in effect, symbols
starting with upper-case letters are variables.

\paragraph{Equality Symbols.}
How do we recognize binary relations as equality relations?  The
default rule is that the symbol \texttt{=} and symbols matching the
pattern [\texttt{Ee}][\texttt{Qq}].* are equality symbols.  If the
input contains the command \texttt{set(tptp\_eq)}, then \texttt{equal} is
the one and only equality symbol.

\paragraph{Infix Notation.}
One can declare binary symbols to be infix and to have a precedence
and associativity so that some parentheses can be omitted.  Many symbols
such as \texttt{=} and \texttt{*} have built-in declarations.

\section{How MACE Works}

The methods used by MACE are described in detail in~\cite{dp-quasi}.
Here is a summary.

For a given domain size, MACE transforms the (first-order) input into
an equivalent propositional problem.  This is possible because,
for a fixed finite domain, the first-order problem is decidable.
The propositional problem is then given to a DPLL
(Davis-Putnam-Loveland-Logeman) procedure.  If satisfiability is
detected, the propositional model is transformed into a first-order
model of the original problem.

Consider the following input file.
\begin{center}
\begin{small}
\begin{boxedverbatim}
list(usable).
  even(a).
  -even(x) | even(s(s(x))).
  -even(s(a)).
end_of_list.
\end{boxedverbatim}
\end{small}
\end{center}
MACE first flattens the clauses into a relational form.
This step involves replacing each $n$-ary function with an $n+1$-ary relation.
MACE's output for this example contains something like

\begin{small}
\begin{verbatim}
    Processing clause: -a(v0) | even(v0).
    Processing clause: -s(v0,v1) | -s(v1,v2) | -even(v0) | even(v2).
    Processing clause: -a(v0) | -s(v0,v1) | -even(v1).
\end{verbatim}
\end{small}
If we ask for models of size 3, MACE generates propositional
clauses corresponding to all instances of the transformed clauses
over the set \{0,1,2\}.  The output also contains the statements

\begin{small}
\begin{verbatim}
    Function s/2 well-defined and closed.
    Function a/1 well-defined and closed.
\end{verbatim}
\end{small}
\noindent
which indicate that MACE has generated
propositional clauses asserting that the new $n+1$-ary relations
are functions.  The DPLL procedure finds a model of the
set of propositional clauses, and the propositional model
is transformed into the following first-order model.

\begin{small}
\begin{verbatim}
    a: 2        even:  0 1 2          s:     0 1 2
                   ---------             ---------
                       T F T                 0 0 1
\end{verbatim}
\end{small}

\paragraph{Scalability.}  Unfortunately, this method does not scale well
as the domain increases or as the size of clauses increases.
Consider a distributivity axiom, $x * (y + z) = (x + y) * (x + z)$.
The transformation to relational form produces the following two clauses.

\begin{small}
\begin{verbatim}
    -+(v0,v1,v2) -+(v0,v3,v4) -*(v4,v2,v5) -+(v3,v1,v6) *(v0,v6,v5)
    -+(v0,v1,v2) -+(v0,v3,v4) *(v4,v2,v5) -+(v3,v1,v6) -*(v0,v6,v5)
\end{verbatim}
\end{small}
\noindent
For a domain of 6, each of these (7-variable) clauses produces $6^7=279,936$
propositional clauses.  MACE can usually handle this many clauses,
but it's hard to fight exponential behavior.  The program SEM~\cite{sem}
is usually better than MACE for large clauses or large domains.

\section{Differences from Previous Versions} \label{history}

Major changes from earlier versions of MACE are listed here.

\begin{enumerate}
\item
Previous versions of MACE called Otter to parse the input
and to produce an intermediate form that was given to a program
named ANLDP.  MACE 2.0 is self-contained, making it easier to install and run.
\item
Previous versions of MACE worked for a fixed domain size, and
there was a separate script (mace-loop) to iterate through
domain sizes and calling MACE.
\item
Previous versions of MACE used Otter's \texttt{passive} list
for constraints (assignments and properties).  MACE 2.0 uses the new
list \texttt{mace\_constraints} for that purpose; clauses in \texttt{passive}
are now taken as part of the theory.
\item
MACE 2.0 allows answer literals in the clauses.  (Answer literals
are removed by MACE before the search for models.)
\item
Previous versions of MACE could handle sorted logic (with
disjoint domains).  MACE 2.0 cannot.  Most of the code for
sorted logic is still in place, so it is possible that
future versions will handle sorted logic.  

Sorted logic can sharply cut down the search time.  Consider
a domain of size 12 that can be partitioned into 8 and 4.
A 2-variable relational clause, with one variable for each sort,
produces 144 propositional clauses with unsorted logic
and 32 clauses in the sorted case.
Let us know if you need sorted logic.
\item
Previous versions of MACE had a checkpointing feature whereby
the state of the search was periodically backed up to a file,
and the search could be resumed from one of those states.
MACE 2.0 does not have this feature.
\end{enumerate}

\section{Calling MACE From Other Programs}

MACE returns an exit code when it terminates.
This makes it convenient to call MACE from other programs.
Here is a list of MACE's exit codes.
(This list changes from time to time; the current
list can be found in the source file \texttt{Mace.h}.)
\begin{description}
\item{\texttt{11 (ABEND\_EXIT)}}
This usually indicates an error in the
input (not all input errors are covered by
\texttt{INPUT\_ERROR\_EXIT} below).
Occasionally it is caused by a bug in MACE.
When you get this exit code, look in the output for an error message.
\item{\texttt{12 (UNSATISFIABLE\_EXIT)}}
MACE completed its search and determined that
no models exist within the given domain size(s) and other constraints.
\emph{It does not mean that the input clauses are unsatisfiable.}
\item{\texttt{13 (MAX\_SECONDS\_EXIT)}}
MACE terminated because of the time limit given on
command line (with \texttt{-t}).
\item{\texttt{14 (MAX\_MEM\_EXIT)}}
MACE terminated because of the memory limit given
on the command line (with \texttt{-k}).
\item{\texttt{15 (MAX\_MODELS\_EXIT)}}
MACE terminated because it found the number of
models requested on the command line (with \texttt{-m}).
\item{\texttt{16 (ALL\_MODELS\_EXIT)}}
MACE completed its search and found all models
(at least one) within the given constraints.
\item{\texttt{17 (SIGINT\_EXIT)}}
MACE terminated because it received the interrupt signal.
\item{\texttt{18 (SEGV\_EXIT)}}
MACE crashed.
\item{\texttt{19 (INPUT\_ERROR\_EXIT)}}
Errors were found in the input.
The output file should point to the error(s).
\end{description}

Say we have a list of equations containing a binary function
symbol \texttt{f}, and we wish to remove the equations
that have a noncommutative model of size $\leq 4$.
If we put the equations in a file, with one equation on
each line, for example,
\begin{center}
\begin{small}
\begin{boxedverbatim}
f(f(x,f(f(z,x),x)),f(z,f(y,x))) = z.
f(f(f(x,f(z,x)),x),f(z,f(y,x))) = z.
f(f(f(f(y,x),z),x),f(f(u,y),x)) = x.
f(f(f(f(y,x),z),x),f(f(y,u),x)) = x.
\end{boxedverbatim}
\end{small}
\end{center}
we can write a simple program to loop through the
equations, calling MACE for each and printing those
that have no noncommutative models of size $\leq 4$.
Here is an example Perl program that does the job.
\begin{center}
\begin{footnotesize}
\begin{boxedverbatim}
#!/usr/local/bin/perl5

$mace = "/home/mccune/bin-linux/mace"; # MACE binary
$unsatisfiable_exit = 12;  # exit code of interest
$input = "/tmp/mace$$";    # temporary input file

while ($equation = <STDIN>) {
    open(FH, ">$input") || die "Cannot open file $input";
    print FH "list(usable). $equation f(0,1)!=f(1,0). end_of_list.\n";
    close(FH);
    $rc = system("$mace -N4 < $input > /dev/null 2> /dev/null");
    $rc = $rc / 256;    # This gets the actual exit code.
    if ($rc == $unsatisfiable_exit) { print $equation; }
}
system("/bin/rm $input");
\end{boxedverbatim}
\end{footnotesize}
\end{center}
If our data file is named \texttt{identities} and our Perl script
is named \texttt{commute4\_filter}, then the command

\begin{small}
\begin{verbatim}
    %
\end{verbatim}
\end{small}
\noindent will remove two of the four equations within a few seconds.

\section{The ANLDP Propositional Decision Procedure} \label{anldp}

If you have a propositional (SAT) problem in conjunctive normal
form, you can call MACE's DPLL procedure directly with the
program ANLDP.
ANLDP is included in the MACE distribution package.

Input to ANLDP is a sequence of integers (no comments are allowed).
The propositional variables are \texttt{1,2,3,\ldots}.
Positive integers are positive literals,
negative integers are negative literals,
and 0 marks the ends of clauses.
For example, here is an (unsatisfiable) input consisting of
four 2-literal clauses.

\begin{small}
\begin{verbatim}
    1 2 0
    1 -2 0
    -1 2 0
    -1 -2 0
\end{verbatim}
\end{small}
\noindent The command-line options of ANLDP are a subset of MACE's:
\begin{clo}
\item[-p]
(Lower case.) This tells ANLDP to print models as they are found.
\item[-m $n$]
This tells ANLDP to stop after finding $n$ models.  The default is 1.
\item[-t $n$]
This tells ANLDP to stop after about $n$ seconds.  The default is unlimited.
\item[-k $n$]
This tells ANLDP to stop if it tries to allocate more than $n$ kilobytes
of memory. The default is 48000 (about 48 megabytes).
\item[-s]
This tells ANLDP to perform unit subsumption as it searches.
(Unit subsumption is always performed on the input.)
When ANLDP gets a new unit (by splitting or by unit propagation),
two operations are ordinarily performed: (1) unit resolution, to
remove complementary literals from all clauses, and (2) unit
subsumption, to mark as subsumed all clauses containing the
unit as a literal.  Because of our data structures, unit subsumption
nearly always costs more time than it saves.  But this option
allows you to use unit subsumption if you wish.
\end{clo}

ANLDP is an implementation of the Davis-Putnam-Loveland-Logeman
procedure.  Efficient data structures and algorithms are used,
but the procedure is otherwise standard.
When the time comes to select the next propositional variable
for splitting, ANLDP simply takes the first variable of the first shortest
positive clause.
Details of the implementation can be found in~\cite{dp-quasi}.

\addcontentsline{toc}{section}{References}
\bibliographystyle{plain}

\bibliography{/home/mccune/papers/bib/master}

\end{document}